# Reliable and Real-Time End-to-End Delivery Protocol in Wireless Sensor Networks

Hossein Pourakbar
Department of Computer Engineering
Islamic Azad University, Tabriz Branch
Tabriz, Iran
husein.msc@gmail.com

Ali Ghaffari
Department of Computer Engineering
Islamic Azad University, Tabriz Branch
Tabriz, Iran
a.ghaffari@iaut.ac.ir

*Abstract*—many routing protocols have been proposed to handle reliability and real-time routing energy efficiency for wireless sensor networks. In this paper we propose a new routing protocol with QoS based capabilities for WSNs. We used priority queues for improve real-time and non-real-time packets forwarding according to deadline of them. The protocol finds a best-cost, time-sensitive packet forwarding mechanism for real-time data with minimum consumption of the energy. In order to avoid of congestion in network our protocol drops those packets who can't reach their destination in specified time. For service quality assurance in reliability domain we used packet reception rate as an important parameter (PRR) in selecting of neighbor nodes. Simulation results show that our new approach how can provide quality of service parameters.

## I. INTRODUCTION

Recent advances in VLSI circuits and contrasting low-power design strategies have led to active research about large-scale, high performance and small-size electronic parts, which are unattended, highly distributed and widely used called sensors. Sensors have many capabilities of detecting environment conditions such as temperature, sound, or the presence of certain objects, and monitoring ambient. In the last few years, due to strategies of minimization and high performance result, design and implementation mechanisms have become more different. For example in a battlefield or a natural disaster, sensors must show their potential to engage of these. The importance of this fact makes us working to increase performance parameters. The objective is to achieve and process important data in nearby or monitoring targets in a specific area. Commonly this data gathered from multiple sensors and become useful information with together. In a jungle flame management setup a large number of sensors can be dropped by a helicopter. Monitoring these mini agents can assist rescue team by locating of fire bottlenecks, identifying the areas have risk of death, and making the rescue operations more aware of the overall event situation. Reliable delivery of data has been one of the challenging areas in wireless sensor network research. It usually studied as important part of network layer [1, 2, 3, and 4] which has multi-hop communications. Respectively, there is a major energy consumption limitation for the sensor nodes. As a resulting, many of new algorithms have been proposed for delivery failure of data problem in wireless sensor networks.

The concentration of studies are mostly on reliability of transfer of sensors is based on protocols which are energy-aware and can maximize the lifetime of all network, scalable for vast number of sensors and immune to battery form running low. In that situation, a service differentiation mechanism is needed in order to guarantee the reliable delivery of the real-time data. Some special type of sensor networks are wireless sensor networks which in sensors interact together in air medium. Consider a security camera field which sends multimedia data to base station periodically. As regards multimedia packets are data-sensitive, thus we must provide such a mechanism to ensure consistency. Our proposed protocol extends the reliability and relaying approach in and considers only end-to-end delay. The protocol looks for a delay-constrained path with the least possible cost based on a cost function defined for each link. Alternative paths with bigger costs are tried until one, which meets the end-to-end delay requirement and maximizes the throughput for best effort traffic is found. Our protocol does not bring any extra overhead to the sensors.

## II. SENSOR NETWORK ARCHITECTURE

In the architecture we consider, sensor nodes are normally distributed in the area sending own data and transmitting others to the sink node. Sensors are only capable of radio-based short-haul communication and are responsible for probing the environment to detect a target/event. Sensors are homogeneous and are same in process, memory, and other parameters, knowing their location and can be aware of others. The source node starts gathering multimedia data such as an image of environment and sends to the next node along the path to the sink. Data packets can be Real-time or non-Real-time. Most of the multimedia data is Real-time, thus when a RT packet is received it must be transmitted immediately then we spot two queues for each relaying node. For avoiding network congestion we defined a deadline for each packet. When a packet's deadline expired, the packet dropped automatically by intermediate node.

### A. Abbreviations and Acronyms
Reliable and Real-time End-to-End Delivery (RREED)
Quality of Service (QoS)
Wireless Sensor Networks (WSN)

### B. Related works
A comprehensive review of the challenges and the real-time communication in sensor networks can be found in [19]. In that paper, the most common WSN routing protocol is presented. A real-time architecture and protocols (RAP) based on velocity can be found in [20]. RAP provides service differentiation in the timeliness domain by velocity-monotonic classification of packets. Based on packet deadline and destination, its required velocity is calculated and its priority is determined in the velocity-monotonic order so that a high velocity packet can be delivered earlier than a low velocity one. Similarly, SPEED is a stateless protocol for real-time communication in WSN. It bounds the end-to-end communication delay by enforcing a uniform communication speed in every hop in the network through a novel combination of feedback control and non-deterministic QoS aware geographic-forwarding [21]. MM-SPEED is an extension to SPEED protocol [22]. It was designed to support multiple communication speeds and provides differentiated reliability. Scheduling messages with deadlines focuses on the problem of providing timeliness guarantees for multi-hop transmissions in a real-time robotic sensor application. In such application, each message is associated with a deadline and may need to traverse multiple hops from the source to the destination. Message deadlines are derived from the validity of the accompanying sensor data and the start time of the consuming task at the destination. The authors propose heuristics for online scheduling of messages with deadline constraints as follow: schedules messages based on their per-hop timeliness constraints, carefully exploit spatial reuse of the wireless channel and explicitly avoid collisions to reduce deadline misses. In traditional best-effort routing throughput and average response time are the main concerns. While many mechanisms have been proposed for routing QoS constrained real-time multimedia data in wire-based networks, they cannot be directly applied to wireless sensor networks due to the limited resources, such as bandwidth and energy that a sensor node has. On the other hand, a number of protocols have been proposed for QoS routing in wireless ad-hoc networks taking the dynamic nature of the network into account [22]. Some of the proposed protocols consider the imprecise state information while determining the routes. In this paper the total waiting time calculation is done hop-by-hop using classic formulas from M/G/1 queues and energy-cost based algorithm. Recent works led to many advances in real-time routing or QoS field but there is no solution which fit them all together, our protocol focused on reliability in delivery and this goes to be realized as a new approach for Real-time systems. Our protocol (called RREED) uses aspects of queuing and bringing the benefits of statistical based decisions satisfying packet reception rate and discusses about energy.

*C. Reliable end-to-end delivery*

Multimedia or real-time content has several unique characteristics as compared with text-based information. First, the size of multimedia content, especially image data, is much greater than that of text-based data. Multimedia data consist of large-sized groups of numerical values, whereas text-based data such as temperature or brightness are usually expressed as a single numerical value. One small image of a place would require about some Kbytes, which in turn requires many packets to transmit because the packet size in a wireless sensor network is usually quite small. Second, multimedia data is quite sensitive to data loss, whereas text-based data is relatively tolerant of data loss. The loss of a small fraction of image data leads to the discarding of the entire image or to a drastic reduction in image quality. Because multimedia content is quite large, it requires many packets that cannot be lost or dropped, if high quality is to be realized. As a consequence, in order to deal with multimedia content, a network protocol should provide end-to-end reliability of packet transmission, in consideration of the characteristics of multimedia content Packets with multimedia content in WSN can be dropped in the case of congestion. Some research efforts on WSNs have focused on congestion control, but this research does not guarantee end-to-end packet delivery. Moreover, in WSNs, packet transmission takes place through the air, a medium in which many errors or losses can occur, so that packet transmission in WSNs is generally regarded as quite unreliable. Some research that has claimed to provide reliable packet transmission usually seeks to reduce the packet loss or error that occurs in node-to-node transmission, not end-to-end.

### III. PRIORITY QUEUES AND M/G/1 MODEL

The M/G/1 queue has exponentially distributed interarrival times and an arbitrary distribution for service times. The increase in generality compared to the M/M/1 queue comes with a price: the M/G/1 queue does not have a general, closed form distribution for the number of jobs in the queue in steady state. It does, however, admit a general solution for the average number of jobs in the queue, and application of Little's Theorem provides the corresponding result for the average time spent in the queue. Collectively, these results are known as the Pollaczek-Khinchin mean value formulae. However, the formulae are valid for any scheduling discipline in which the server is busy if the queue is non-empty, no job departs the queue before completing service, and the order of service is not dependent on knowledge about job service times. Job i refers to the ith job to arrive at the queue. As usual, we assume that a steady state solution exists.

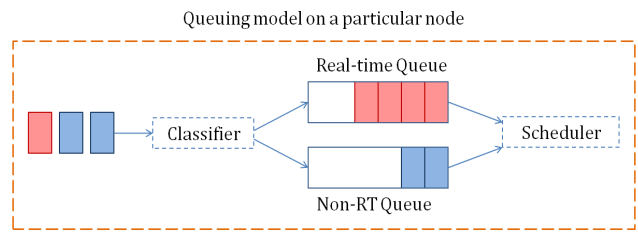

Fig. 1: M/G/1 queuing model

When a job arrives at the queue, it must wait for the job in service (if there is one) to complete. Since we are assuming FCFS scheduling, it must also wait for all of the jobs which arrived before it but which have not begun service to complete.

- $X_i - r.v. \text{ for service time of job } i$
- $\overline{X} = E[X_i]$
- $W_i - r.v. \text{ for time job } i \text{ spends waiting in queue before begining service}$
- $W = \lim_{i \to \infty} \overline{W_i}$
- $N_i - r.v. \text{ for number of jobs in the queue when job } i \text{ arrives}$
- $N_Q = \lim_{i \to \infty} E[N_i]$
- $R_i - r.v. \text{ for residual service time seen by job } i$
- $R = \lim_{i \to \infty} E[R_i]$
- $\lambda - \text{job arrival rate}$
- $\mu - \text{service rate}$

(1)

We can take limit of both sides of this equation to obtain Pollaczek-Khinchin (P-K) formula, At last from little's theorem we have:

$$W_i = R_i + \sum_{j=i-N_i}^{i-1} X_j$$

$$W = R + \overline{X} N_Q, \quad N_Q = \lambda W$$

$$W_1 = \frac{R}{1-\rho_1}, \quad W_2 = R + \frac{1}{\mu_1} N_{Q_1} + \frac{1}{\mu_2} N_{Q_2} + \frac{1}{\mu_1} \lambda_1 W_2$$

$$W_2 = \frac{R}{(1-\rho_1)(1-\rho_1-\rho_2)} \quad (2)$$

## IV. ENERGY COST CALCULATIONS

The amount of energy needed for sending a packet is shown in Fig. 3; we must minimize the size of packet we need to send data in order to minimize energy cost in real world. This energy is calculated for one hop, whereas our protocol is multi-hop. Then we must find a formula to calculate total cost of transmission path along sink node.

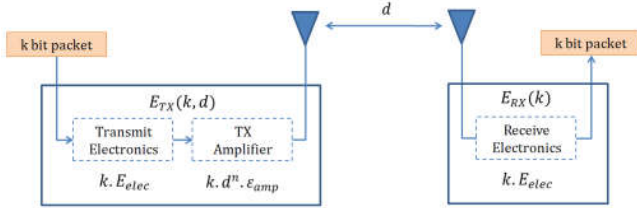

Fig. 3: Radio energy dissipation model

Figure above shows two sides of a trip, Left for transmitter and the right one for receiver. When a packet needs to be sent, the transmitter calculates length of packet in bits and then starts sending it to the proper destination receiver.

$$E_{TX}(k,d) = E_{TX-elec}(k) + E_{TX-amp}(k,d) = \begin{cases} k.E_{elec} + k.\varepsilon_{fs}d^2, & d < d_0 \\ k.E_{elec} + k.\varepsilon_{amp}d^4, & d > d_0 \end{cases}$$

$$\text{where } d_0 \text{ is: } \sqrt{\frac{\varepsilon_{fs}}{\varepsilon_{amp}}} \quad (3)$$

As well, to receive a k-bit message, the radio expends

$$E_{TX}(k) = E_{TX-elec}(k) = k.E_{elec} \quad (4)$$

## V. LINK ESTIMATE WITH PRR

Link estimation is a critical part of almost every sensor network routing protocol. Knowing the packet reception rate of candidate neighbors lets a protocol choose the most energy efficient next routing hop. The earliest sensor link estimators assumed link symmetry, establishing routes either through flooding [5] or snooping [7]. Later it was shown that this assumption was invalid and it led to terrible performance. Second generation estimators [13] used packet sequence numbers to count lost packets, but required a low rate of control traffic to ensure that nodes could detect when links died. While this approach can detect good links, it adapts slowly to changes in link quality. This limitation led to a number of approaches that use information from the radio hardware. We assume there is a parameter PRR [Packet Reception Rate] which indicated rate of successful receiving of packets in a period of time for a specific sensor. Many formulas obtained in related works, but we accept this formula to ensure next node reception rate

$$PRR = \frac{Number\ of\ received\ packets}{Number\ of\ sent\ packets} \quad (5)$$

## VI. CALCULATION OF PATH END-TO-END DELAY

We provide a cost function for each transmission; this cost is consisting of all three parameters: Queue Delay, Residual Energy and Packet Reception Rate. In addition we must select the nearest node to us and destination (sink).

$$Cost_{i,j} = \alpha(delay_i) + \beta\left(\frac{1}{energy_j}\right) + \gamma\left(\frac{1}{PRR_j}\right) \quad (6)$$

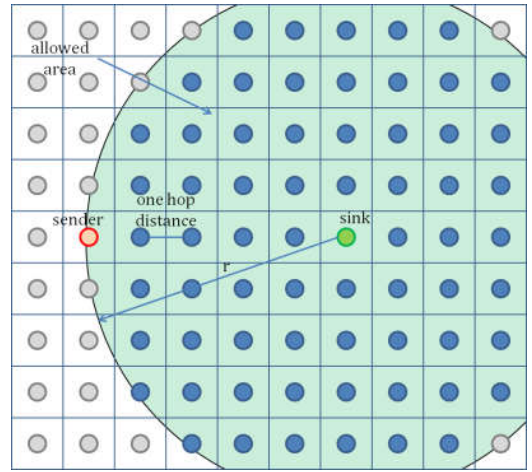

Fig. 4: Sender and neighbors in allowed area

When the packet i needs to be send, source sensor creates neighbors table and calculates the cost function for each of them. Neighbors table build with a mathematical operation, sensor i sends it's coordinates to sink, sink finds all sensors potentially can be neighbor of sensor i, only those sensors cab be neighbor of sensor i which are in a circle with radius of sensor-sink and center of sink oval. This can prevent neighbors of wrong paths. Fig 4 shows how allowed neighbors identified.

We provide a simple proximate to number of hops to sink node, which calculates area of allowed neighbors and then number of allowed neighbors in linear path to sink node. The number $\Delta$ can be used as distance of each node to its neighbor. With sensor distance to sink we have.

$$\Delta = \sqrt[2]{\frac{A_{an}}{N_n}}$$

where $A_{an}$ is: Area of allowed neighbors (shown in green color)

and $N_n$ is: Number of neighbors in $A_{an}$ (blue ovals)

$$N_{LPS} = \frac{dist_{sensor-sink}}{\Delta}$$

where $N_{LPS}$ is: number of hops in linear path (7)

## VII. ALGORITHMS AND EXPERIMENTAL RESULTS

A packet with a specific deadline can wait in a short period of time along its path. Then calculating delay for each node results to obtain path end-to-end delay. If we know each path approximately delay and energy capabilities of next neighbor we can guarantee respective reliability to the normal state. As we said before, there are two queues in receiver's memory, one classifier and a scheduler, when a packet arrive into classifier it must enqueue it in the proper queue based on RT or non-RT type of packet. Then packet must wait for its service time to be come. The algorithm provided below can simulate the receiver node in steady-state.

```
Algorithm used to transfer a packet (single hop)
• Calculate delay parameter
1. Set Δ = Sqrt (A_an/N_n)
2. Set N_LPS = dist_sensor-sink/ Δ
3. NT = Neighbors_table(single hop)
4. For each N in NT
5.     If packet is Real-time
6.         Set delay_i = R/(1-ρ_1)
7.     Else
8.         Set delay_i = R/((1-ρ_1)(1-ρ_2))
9. Next N

• Calculate energy parameter
1. k = number_of_bits
2. For each N in NT
3.     Set energy_j = E_residual_j − k.E_elec
4. Next N

• Find best neighbor
1. Init α, β, γ
2. Set cost_min = ∞
3. For each N in NT
4.     Set cost_i,j = α(delay_i) + β(1/energy_j) + γ(1/PRR_j)
5.     If cost_min > cost_i,j
6.         Set cost_min = cost_i,j
7. Next N
8. Return cost_min and N_min as result
```

```
Algorithm used in receiver node
1. If packet.type = RT
2.     Enqueue_RT(packet)
3. Else
4.     Enqueue_NoneRT(packet)
5. If X_j in scheduler = 0
6.     Set p = Dequeue()
7.     Schedule(p)
8. Else
9.     X_j = X_j − 1

1. Do
2.     Scheduler.fetch_next()
3. While there_is_packets
```

## CONCLUSION AND FUTURE WORK

In this paper, we presented a new reliable real-time protocol for multimedia data in sensor networks field. The protocol finds better paths for real-time data with certain end-to-end delay requirements. The effectiveness of the protocol is validated by simulation. Simulation results show that our protocol consistently performs well with respect to reliability metrics, e.g. throughput and average delay as well as energy-based metric such as average lifetime of a node or the entire network's body. The results have also shown that real-time data rate, buffer size, and packet drop probability have significant effects on the performance of the protocol. We are currently extending the model and trying to minimize the path delay and plan to compare the performance of such extended model with the Reliable Real-time End-to-End transfer protocol for multimedia data presented in this paper. The graphical diagram result attached in appendix cab show the effectiveness of this protocol, we will focus on better results later. These outputs are results of several simulations. The parameters used for simulation is shown below.

| Parameter | Value |
|---|---|
| E(init) | 2J/batt |
| Data Packet size | 100 bits |
| E(elec) | 50 nJ/bit |
| E(amp) | 0.0013 pJ/bit/m4 |
| E(fs) | 10 pJ/bit/m2 |
| Network Grid | (0, 0) to (1000, 1000) |
| α | 0.6 |
| β | 0.3 |
| γ | 0.1 |

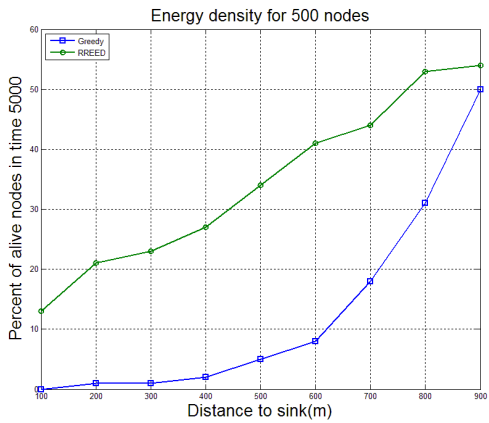

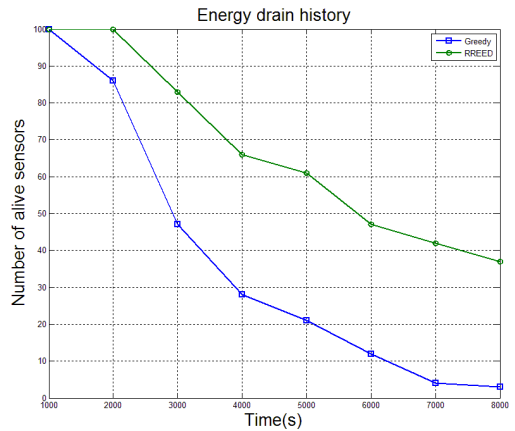

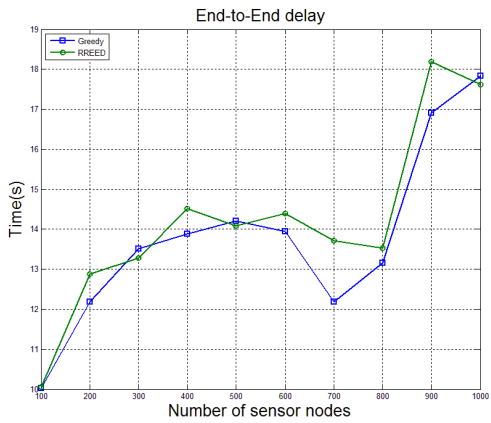

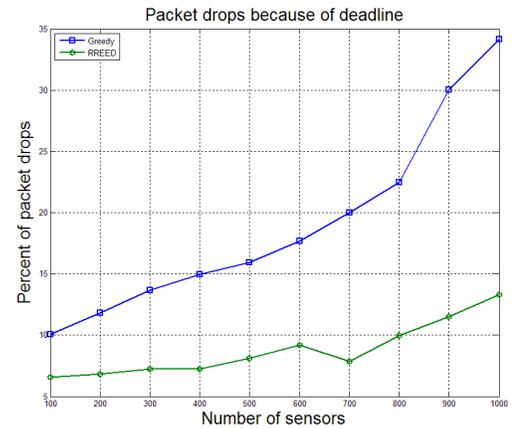